\documentclass[preprint,proceedings]{rmaa}

\suppressfulladdresses

\usepackage{paralist}
\usepackage{psfrag,color}
\SetYear{2007}
\SetConfTitle{Binary Systems...}

\title{The extraordinary LBV$/$WR system HD 5980}

\author{Gloria Koenigsberger \altaffilmark{1}
        and Edmundo Moreno\altaffilmark{2}
       }
\altaffiltext{1}{Instituto de Ciencias F\'{\i}sicas, Universidad Nacional Aut\'onoma de M\'exico, 
Apdo. Postal 48-3, Cuernavaca, Morelos 62251, M\'exico (gloria@ce.fis.unam.mx).}

\altaffiltext{2}{Instituto de Astronom\'{\i}a, Universidad Nacional Aut\'onoma de M\'exico,
Apdo. Postal 70-264, M\'exico D.~F., 04510, M\'exico (edmundo@ce.fis.unam.mx).}

\shortauthor{Koenigsberger \& Moreno}
\shorttitle{Tides in HD5980}
\fulladdresses{
\item Gloria Koenigsberger: Instituto de Ciencias F\'{\i}sicas, Universidad Nacional Aut\'onoma de M\'exico,
Apdo. Postal 48-3, Cuernavaca, Mor. 62251, gloria@ce.fis.unam.mx
\item Edmundo Moreno: Instituto de Astronom\'{\i}a, Universidad Nacional Aut\'onoma de M\'exico,
Apdo. Postal 70-264, M\'exico D.~F., 04510, edmundo@ce.fis.unam.mx
}

\listofauthors{G. Koenigsberger, E. Moreno}
\indexauthor{Koenigsberger, G.}
\indexauthor{Moreno, E.}

\resumen{ Presentamos resultados de los c\'alculos para HD 5980 efectuados con nuestro modelo
de interacci\'on por fuerzas de marea.  El modelo predice una mayor actividad
atmosf\'erica despu\'es del paso por el periastro, as\'{\i} como una mayor
actividad ocasional en zonas intermedias entre el ecuador y el polo. 
}

\abstract{The LBV/WR system HD 5980 contains a short-period, eccentric binary system
with interacting stellar winds.  In this paper we  summarize results from model
calculations of the tidal flows on the LBV component showing that  energy 
dissipation rates, $\dot{E}$, associated with turbulent viscosity are orbital-phase 
dependent as well as variable over the stellar surface. We speculate that if $\dot{E}$ 
contributes towards  driving  mass-loss, the strongest wind-wind interaction effects may  
occur {\em after} periastron passage.  In addition, the model suggests the presence of 
stronger outflows localized at polar angles $\theta \sim$30 --50$^\circ$  
during part of the orbital cycle.   Thus, the analysis of wind-wind
interactions in this system requires that models be revised to incorporate  non-stationary and 
asymmetric wind structures.
}

\addkeyword{Binaries: close}
\addkeyword{Binaries: tidal effects}
\addkeyword{Binaries: interacting winds}
\addkeyword{Stars: individual (HD5980)}

\begin{document}
\maketitle

\section{Eruptions, wind-wind interactions and tides}

HD 5980 is an amazing system. Niemela (1988) was the first to point out that it consists of 
two Wolf-Rayet-like  components referred to as {\it star A} and {\it star B} in a relatively 
close, eclipsing and excentric orbit (P$=$19.3, e$\sim$0.3) and a third source, referred to as 
{\it star C}, that may simply lie along the line-of-sight to the close pair. The spectral
characteristics and visual brightness of the system  underwent significant changes
between the late 1970's and 1993, when it entered an eruptive state that lasted
$\sim$1 year (Bateson \& Jones 1994; Barb\'a et al. 1995; Koenigsberger et al. 1995).
The  activity  involved an increase in visual brightness and mass-loss rate, and a 
decrease in  wind velocity  and  effective temperature, similar to the eruption
phenomena observed in luminous blue variables (LBVs).  The radial velocity variations observed
in the very rich emission-line spectrum that appeared after the eruption led Barb\'a et al. 
(1996; 1997)  to conclude that the instability producing the outbursts originated in star A. 
A detailed review of HD 5980's properties is provided by Koenigsberger (2004).

The ZAMS masses of star A and star B are inferred to be $\geq$100 M$_\odot$ (Koenigsberger 2004).  
The substantial mass  loss required  for them to have reached their current masses 
(M$_A\sim$50 M$_\odot$, M$_B\sim$30 M$_\odot$, Niemela et al. 1997) may have been achieved
through multiple events as those of 1993/1994.  LBV's are associated with the 
evolutionary state during which large quantities of mass are ejected  allowing the star to 
reach the W-R  phases with highly depleted hydrogen envelopes.  With the possible exception
of $\eta$ Carinae, there is no known LBV with such a developed W-R spectrum as that displayed
by HD 5980. 

\subsection{A changing wind-momentum ratio}
The spectrum of HD 5980 in the late 1970's, with its broad He II and N V lines,  
was typical of the ``early" W-R stars of the nitrogen sequence (WNE; van der Hucht, 2001). 
This spectrum is believed to originate in the  wind of star B. Over the next decade, 
however, numerous lines from lower-ionization atomic species  appeared, implying a growing 
presence of a cooler stellar wind, which we now attribute to star A.  Clearly, the 
emerging dominance of star A's wind implies changes in  the wind-wind interaction (WWI) 
region characteristics,

Emission-line profile variations observed in optical and UV wavebands are phase-locked and 
should, in principle, provide information on the geometry of the changing WWI region
(Moffat et al. 1998). Surprisingly, however, the nature of the variability has remained 
the same ever since it was first reported by Breysacher \& Westerlund (1978) and quantified by 
Breysacher, Moffat \& Niemela (1982). The variations consist of periodic changes in width 
and degree of asymmetry as a function of orbital phase.   Figure 1 illustrates the 
NIV] 1486 \AA\ emission line variability: it is always narrower and sharply peaked  
near the eclipse when star B is ``in front" ($\phi\sim$0.40), while becoming broader and 
weaker when both stars are unocculted at $\phi=$0.83.  The three epochs that are displayed  
correspond to pre-eruption (1991), post-eruption (1999) and $\sim$ 1 year after maximum (1995). 
This persistent trend is  unexpected since the geometry of the WWI regions depends on the momentum 
ratio of the stellar winds, a ratio  that changed over time as star A's wind became 
more dominant.  Thus,  HD 5980 appears to provide another  example of the discrepancies that 
arise when confronting the current WWI models with observations (see presentations
in this Volume by Rauw, Pollock, Williams, among others).   But at the same time, 
because of the  large observational database available for  HD 5980, its behavior 
may provide a clue to identifying   the source of the discrepancies. 

\begin{figure}[!t]
\includegraphics[width=\columnwidth]{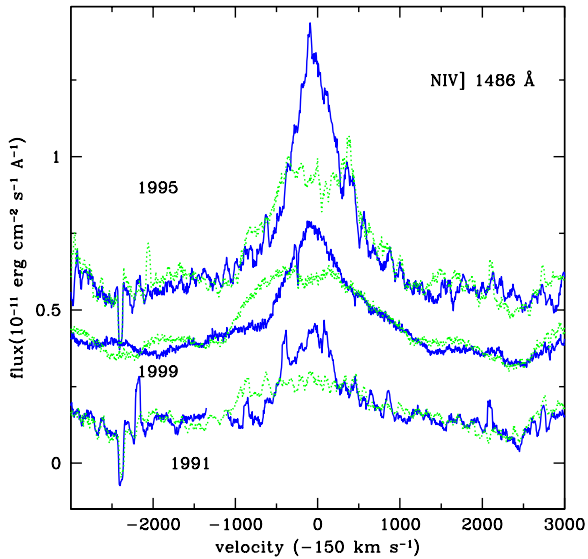}
\caption{Line profiles of the semi-forbidden N IV emission line at $\lambda$ 1486 \AA~
observed at two different orbital phases ($\phi\sim$0.40 and 0.83 --dots) for each of
three different epochs (1991, 1995 and 1999).  The qualitative nature of the line-profile
variations remains constant. Sets of profiles for different epochs are vertically displaced
for clarity in the figure.}
\end{figure}

\subsection{Tides and non-stationary, asymmetric winds}

Like many of the intriguingly active binaries, HD 5980 has an excentric orbit.  In such 
systems, the tidal forces are time-variable.  The preliminary exploration  of
tidal effects in HD 5980 led to the conclusion that they are non-negligible (Koenigsberger 
et al. 2002), and thus raised  the question of whether they may be responsible for 
some of the system's peculiarities.  
Recent results of our calculations (Moreno \& Koenigsberger 2007) indicate that the tidal flows 
near the stellar surface can liberate considerable amounts of energy through dissipative 
processes. The magnitude of the energy dissipation rate, $\dot{E}$, depends on the stellar
and orbital parameters.  In the following sections, we'll describe the two basic conclusions
of the calculations: 1) maximum $\dot{E}$ occurs {\em after} periastron passage, not at periastron; 
and  2) at certain orbital phases,  larger values of $\dot{E}$ are generated at intermediate 
polar angles than in the  equatorial belt.  These results are relevant for WWI theory
if we assume that $\dot{E}$ contributes towards enhancing the stellar  mass-loss rate.
The winds in excentric binaries such as HD 5980 would then be {\em intrinsically} 
non-spherically symmetric and time-dependent,  thus leading to discrepancies  when comparing 
the observational diagnostics of WWI with the predictions of stationary models.

\section{Energy dissipation from tidal flows in HD 5980}

Tidal effects are important when a star's rotation, $\omega$, is not synchronized with the
orbital motion, $\Omega$.  In excentric binaries, this is generally the case since the orbital
motion varies with phase while the stellar rotation rate remains constant.  There is no
direct observation of $v sin i$ for star A or star B. But the low-amplitude variations of 
star C's narrow absorption lines over the 19.3-day orbit has been interpreted
in terms of contamination by very broad absorptions arising in star A but not visible due
to their superposition on the  emission lines (Georgiev \& Koenigsberger 2004).  Under 
this assumption,  $v sin i\geq$200 km/s, thus providing an estimate for the ratio 
$\omega$/$\Omega_{per}\sim$2.33, where $\Omega_{per}$ is the orbital angular velocity at
periastron.  This means that star A rotates super-synchronously throughout the orbital
cycle.  

The  basic method  is described in Moreno \& Koenigsberger (1999) and Moreno et al. (2003).
With the recent extensions (Moreno \& Koenigsberger, 2007), the code now computes the amplitudes of
the tidal flow in a thin surface layer  over the entire stellar surface.  These amplitudes
are used to estimate  the shear energy dissipation rates, $\dot{E}$  that arise from the 
relative motions of different surface layers  using an extension of the approach described 
in Toledano et al. (2006).

\begin{figure}[!t]
\includegraphics[width=\columnwidth]{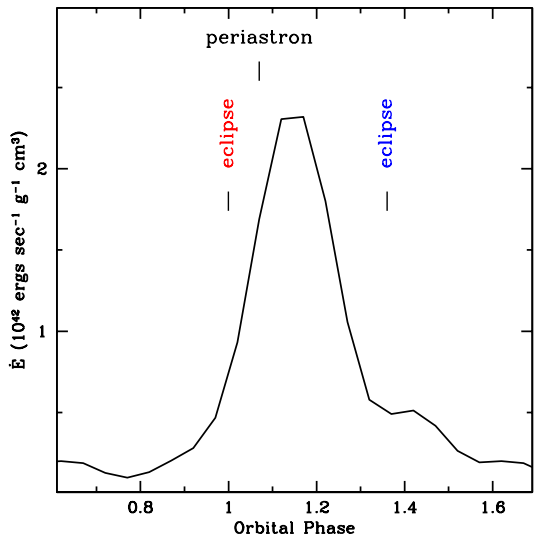}
\caption{Predicted energy dissipation rate per unit density due to the tidal
flows on the stellar surface of star A as a function of orbital phase. Maximum
rates occur {\em after} periastron passage.  Star A is ``in front" at $\phi=$0,
and periastron passage occurs at $\phi \sim$0.07.
}
\end{figure}

\begin{figure}[!t]
\includegraphics[width=\columnwidth]{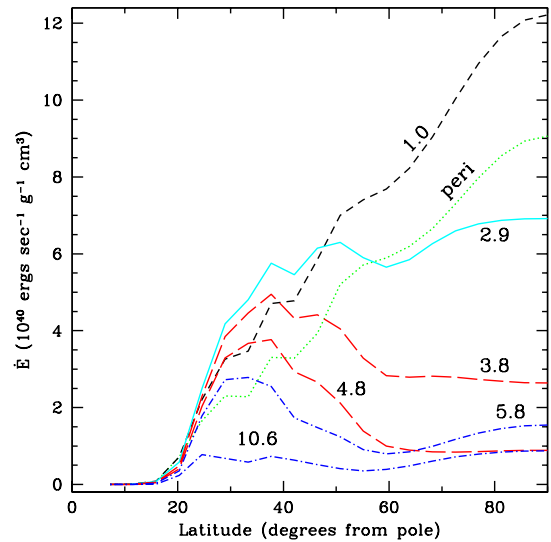}
\caption{Predicted energy dissipation rate per unit density due to the tidal
flows on the stellar surface of star A as a function of latitude at periastron and
several times (in days) after periastron passage. Note that $\dot{E}$ is greater  
at intermetidate polar angles ($\theta \sim$20--50 $^\circ$) than at the equator
around $\sim$days 3--6 after periastron.
}
\end{figure}

Figure 2 illustrates the time-dependence of $\dot{E}$ as a function of orbital
phase from our HD 5980 model calculation.  The first thing to note is that the 
maximum rates are generated after periastron passage, $\phi \sim$0.1--0.25, with
minimum values around $\phi=$0.8.  Thus, it is tempting to suggest that the
persistent line-profile variability shown in Figure 1 may be associated with
the orbital-phase dependent changes in $\dot{E}$.   But in what way do the  tidal 
effects produce these variations ?

The answer may lie in  the  distribution in latitude  of  $\dot{E}$. Figure 3 illustrates 
$\dot{E}=\dot{E}(\theta)$ for several different times within the orbital cycle.   
At periastron (dotted curve) and until $\sim$2 days thereafter, maximum $\dot{E}(\theta)$ 
occurs at the equator  and systematically decreases towards the pole ($\theta=$0 at the pole).   
By day $\sim$3 and until apastron, however, there is a distinct change in this trend  whereby 
maximum $\dot{E}(\theta)$ now occurs at $\theta \sim$30--50$^\circ$.   If we assume that 
$\dot{E} \rightarrow \dot{M}$, the results shown in Figure 3 imply that between days $\sim$3 
and $\sim$6  after periastron passage ($\phi \sim$0.2 -- 0.4) mass outflow  from regions at  
intermediate polar angles is more intense than from the equator.  A denser ``polar" wind 
would produce a  narrower and sharply peaked emission-line profile, as seen in Figure 1 for
$\phi \sim$0.40.  Additional observational evidence for polar  outflows within this orbital 
phase interval is provided by  Villar-Sbaffi et al. (2003) who state that ``the mass-loss of HD 5980 
around $\phi=$0.36 presented fluctuations in axial symmetry ranging from very rapid density  
enhancements along the orbital plane to polar ejections."   At orbital phase $\phi\sim$0.8, 
$\dot{E}(\theta)$ has a relatively small gradient between the pole and the equator and is 
significantly weaker than near periastron.  Thus, the stellar wind structure of star A should 
be more spherically symmetric at this phase. 

Clearly, there are additional contributions to producing the line-profile variations, such as
the physical and wind eclipses, as well as contributions from the WWI zone and all of these
need to be considered.  However, it is encouraging that the asymmetries in wind structure
predicted by the  tidal interaction model are consistent with the emission-line profile
variability.

\section{Final reflections}

A one-layer tidal interaction model for HD 5980  predicts that the energy dissipation 
rate due to the tidal shearing flows are time-dependent and non-spherically symmetric.  
Speculating that $\dot{E}$ contributes towards the stellar wind mass-loss, leads to the 
conclusion that $\dot{M}$ may also be locally enhanced near specific surface locations.  
In particular, outflows  at intermediate polar angles  may  be stronger than at the 
equator at particular orbital phases.  Furthermore, the model predicts that maximum $\dot{E}$  
should occur after periastron passage, thus implying an overall enhanced $\dot{M}$ compared 
to other orbital phases.  

Post-periastron events seem to occur in a wide variety of binary systems, such
as WR 140, $\eta$ Carinae and others, raising the question of whether the tidal
effects described above are more prevalent than one may have anticipated.  Is it
possible that the stronger WWI effects  (``outbursts") that occur after periastron passage are 
associated with stronger mass-loss rates at these phases induced by tidal instabilities ?  If
this were the case,  the source of  the discrepancies between wind-wind interaction model 
predictions and  the observations may simply reside in the assumption of stationary and 
spherically symmetric  winds.

It is interesting to note that the hypothesis of enhanced $\dot{M}$ arising
from $\dot{E}$ may not be entirely unreasonable.  Given HD 5980's huge UV luminosity
(see, for example, Koenigsberger 2004), it is likely to be on the verge of the Eddington
Limit, as other W-R stars appear to be (Goeffner, this workshop).  Thus,  small additions
of energy in sub-photospheric layers could drive it to a super-Eddington state. We speculate
that viscous shear energy dissipation resulting from the tidal forces  may be a non-negligible
contributor to this small needed additional energy.

Within this context, a final consideration concerns the sudden eruptive events in HD 5980.  
Monitoring of HD 5980 at visual and UV wavebands prior to, during and after its eruptions has
yielded a unique data set  that provide clues for constraining the eruption mechanism.  For
example, ultraviolet observations  suggest that the onset of the eruptive state involved rapid
transitions between a fast and a slow stellar wind (Koenigsberger 2004).  Hence, it is likely
that the eruption occurred when  the wind became so dense that the bistability limit was 
crossed (Lamers, Snow \& Lindholm 1995). But this is only the symptom of a more deep-seated
phenomenon that causes the instability leading to the enhanced density wind in the first place. 
Tidal effects are very sensitive to the star's radius.  If HD 5980 (and other similar stars) are 
undergoing an evolutionary transition by which outer layers are expanding, the amplitudes of 
the tidal flows are expected to grow significantly.  If the hypothesis that 
$\dot{E} \rightarrow \dot{M}$ can be shown to stand on firm ground,  this would provide a
mechanism to remove significant amounts of mass  as the star tries to evolve towards the 
red end of the H-R Diagram.  Whether the mass-shedding occurs as episodic eruptions, such 
as we've observed in HD 5980, or whether it is through a sustained high-density wind, requires 
an understanding of the $\dot{E} \rightarrow \dot{M}$ process.  Since our model neglects the 
effects of intrinsic stellar oscillation modes,  effective temperature variations and radiation 
pressure,  we are unable to go beyond the  speculative realm at this time.

\begin{acknowledgements}

GK thanks Jeff Kuhn and Stan Owocki for very helpful discussions. 
Support from PAPIIT/DGAPA grant IN119205 is gratefully acknowledged.

\end{acknowledgements}

{}{}
\clearpage

\clearpage
\vfill\eject

\centerline{Transcription of questions and answers}

\noindent {\bf Moffat:} There are other sources of perturbation that should be considered, 
such as the heating effects from the X-rays produced at the bow-shock head between stars A and B,
as we started to do in Moffat et al. (1998).

\noindent {\bf Reply:}  Wind-wind collisions and the effects associated with them are inevitable
if the winds of both stars are able to accelerate to large speeds in the intervening
region.  However, strong tidal perturbations are also inevitable if the stellar rotation is
not synchronized with the orbital angular velocity.

\noindent{\bf Nathan Smith:} The high luminosity high mass loss make LBV ourbrust and the fact that
its in a binary naturally make me think of comparing HD 5980 to Eta Car.  In that case,
its curious that their outbursts were so different.  Namely, Eta Car ejected over 10 Mo in
its 19th century outburst, whereas HD 5980 only ejected about 0.001 Mo ...that's four orders
of magnitude different and it begs for an explanation, given that the primary star's
luminosities and mass loss rates are within a factor of 2.  One obvious difference that
comes to mind is that Eta Car has a 5 year period whereas HD 5980 has amuch smaller
separation with only a 19.3 day period.  So I am wondering if the closer companion in
HD 5980 somehow regulates the LBV instability so that it erupts more often with less
mass, and therefore doesn't build up the catastrophic instability that leads to a major
outburst like Eta Car.  Any thoughts on that ?

\noindent {\bf Reply:}  We do not know what the underlying mechanism for the instability in HD 5980
really is.  If we assume that its radius is growing due to evolutionary processes,
it is possible that the presence of the binary companion produces a ``premature"
eruption through the tidal oscillations mechanism, removing mass, and thus slowing
(momentarily) the expansion.  The closer the binary companion is, the more frequent
such events would be expected to occur.

\noindent{\bf Moffat:}  (After N. Smith comparing HD5980 to Eta): Eta at periastron could be more
similar to HD 5980, given its large eccentricity.

\noindent {\bf Reply:} The orbital separation in HD 5980 is only $\sim$100 Ro.

\noindent{\bf Maeder:} What about the chemical abundances which may tell us something 
on the evolutionary stage of HD 5980 ?

\noindent {\bf Reply:} The Non-LTE analysis that we made on the eruptor's spectrum (Koenigsberger 
et al. 1998) indicates that there is a significant fraction of hydrogen.  A determination
of other chemical element abundances would be highly desirable.

\noindent{\bf Walborn:} Could the star be hitting the Eddington Limit as well as or instead of the
Bistability Limit as it tries to evolve ?   If there are internal magnetic fields combined with 
differential rotation and tidal oscillations, the effects may also be catastrophic.

\noindent {\bf Reply:} Yes, indeed.

\noindent{\bf Stan Owocki:} (replying to Nolan): One idea for a way to trigger a 
super-Eddington luminosity is that the tidal excitation of pulsations you mentioned is 
associated with a breaking of the spherical symmetry that is essential for blocking the 
radiation in the envelope.  The associated clumping or "porosity" of the envelope could  
then allow radiation to escape from  an edge, leading to super-Eddington  brightening that 
could then drive the mass eruption.

\noindent{\bf Nathan Smith:} (after Stan's comment):  Did the bolometric luminosity of HD 5980 change
during its LBV outburst ? Pinning that down is important for explaining the differences
between this event and Eta Car.   If HD 5980 stays at roughly constant L$_{bol}$, the
likely implication is that it is crossing the bistability jump and develops a
pseudophotosphere as in a normal S Dor outburst, whereas Eta Car violated the classical
Eddington limit and therefore suffered a much more violent, deep seated event.
In that context, one question is does HD 5980 show any evidence for a massive circumstellar
shell ejected in an ancient EtaCar-like eruption ?

\noindent {\bf Reply:}  We (Koenigsberger et al. 1998, ApJ 499, 889) derived 
L$_{bol}=$3$\times$10$^6$ Lo, but Drissen et al. (2001, ApJ 545, 484) derived $\sim$10$^7$, so 
it is not clear whether there was a change or not.

\noindent{\bf Moffat:} Tidal oscillations may indeed be an important trigger of the eruption of
star A.  But once star A reaches maximum size, it is much larger than the
orbital separation.  This can lead to common envelope evolution for a time,
and eventually shorten the orbital period leading in the (very) short period WR+O
binaries we see in the Magellanic Clouds.  This means that the LBV phase may indeed
be crucial for explaining WR stars, given the reduced mass-loss rates of their
progenitor O-stars.

\noindent {\bf Reply:} The common envelope phase (if we can call it that!) during the 1994 
eruption lasted less than a year, probably too short to cause any changes in the orbit. 
But if this were to happen frequently enough...

\end{document}